\documentclass[prl,aps,twocolumn,showpacs]{revtex4}
\usepackage{bm}
\usepackage{amssymb}
\usepackage{graphicx}

\begin{document}

\title{Magnon Decay in Noncollinear Quantum Antiferromagnets}

\author{A. L. Chernyshev}
\affiliation{
Department of Physics, University of California, Irvine, 
California 92697, USA}
\author{M. E. Zhitomirsky} 
\affiliation{
Commissariat \`a l'Energie Atomique, DSM/DRFMC/SPSMS, 38054 Grenoble,
France}

\date{August 3, 2006}

\begin{abstract}
Instability of the excitation spectrum of an ordered noncollinear
Heisenberg antiferromagnet (AF) with respect to spontaneous 
two-magnon decays is investigated.
We use a spin-1/2 AF on a triangular lattice as an example
and examine the characteristic long- and 
short wave-length features of its zero-temperature
spectrum within the $1/S$-approximation.  
The kinematic conditions are shown to be crucial for the 
existence of decays and for overall properties of the spectrum. 
The $XXZ$ and the $J$--$J'$ generalizations of the model, as well as 
the role of higher-order corrections are discussed.
\end{abstract}

\maketitle

A quantum many-body system with nonconserved number of particles may
have cubic vertices, which describe interaction between one- and 
two-particle states. 
In crystals such anharmonicities
lead to finite thermal conductivity by phonons \cite{ziman}. 
In the superfluid $^4$He,   
the cubic interaction between quasiparticles result in a 
complete wipeout of the single-particle branch at
energies larger than twice the roton energy \cite{pitaevskii}.

In quantum magnets with {\it collinear} spin configuration,
{\it e.g.}, in AF on a bipartite lattice, cubic terms are
absent and anharmonicities are of higher order \cite{collinear}.
The cubic terms can appear due to
dipolar interactions \cite{akhiezer}, but in magnetic insulators 
those are weak and usually can be neglected.
It has been gradually realized that substantial 
cubic interactions should exist in the {\it noncollinear} 
AFs \cite{miyake,ohyama,chubukov94,Zh_nikuni}. Qualitatively, 
such cubic anharmonic terms arise due to coupling of the
transverse (one-magnon) and the longitudinal (two-magnon) fluctuations in
these systems.

The noncollinearity of an antiferromagnetic spin configuration can be
either induced by external magnetic field \cite{Zh_nikuni} 
or by frustrating effect of the lattice ({\it e.g.}, in
the triangular lattice (TL) spins form the so-called 
120$^\circ$ structure \cite{miyake,chubukov94}). 
In the former case, spontaneous decays are allowed above a
threshold field $H^*$, such that magnons become strongly damped 
throughout the Brillouin zone (BZ) \cite{field}. 
On the other hand, the role of magnon interactions in
the spectra of frustrated AFs is not well understood. 
The earlier work on TLAF \cite{chubukov94}
has discussed only renormalization of the spin-wave velocities.
The recent series expansion study \cite{Singh} has found a
substantial deviation of the spectrum from the linear spin-wave theory 
(LSWT) and interpreted it as a sign of spinons. 
The latest work \cite{Starykh} has questioned this hypothesis
by showing that $1/S$ expansion strongly modifies the LSWT
spectrum leading to an overall agreement with the numerical data.
However, the subject of spontaneous decays has 
been hardly touched upon. Astonishingly, instability of the 
single-particle spectrum in the presence of a well-defined, magnetically 
ordered ground state is, perhaps, the single most striking qualitative
difference of the non-collinear AFs from the collinear ones.

In this Letter, we shall study magnon decay
in noncollinear quantum AFs at $T=0$ using an example
of the Heisenberg AF on a TL. 
The noncollinearity is necessary but not sufficient for 
 decays. In addition, the energy and momentum must be conserved
within a decay process (kinematic conditions).
Thus, the decays are determined, in part, by the shape of 
the single-particle dispersion that may, or may not, allow 
spontaneous decays. 
We analyze singularities in the
two-magnon continuum that outline instability regions or lead 
to discontinuities in the single-particle spectrum.
Our long wave-length analysis  yields definite asymptotic
statements regarding the life-time of magnetic excitations.
We discuss briefly the $XXZ$ and $J$--$J'$ models where
noncollinearity is also present.

We begin by rewriting the spin-$S$, nearest-neighbor 
Heisenberg Hamiltonian for the TLAF 
into the local rotating frame associated with the classical 
120$^\circ$ structure of the spins, and proceed with the standard
Holstein-Primakoff transformation of the spin operators into bosons 
followed by the Bogolyubov transformation diagonalizing the harmonic part
of the bosonic Hamiltonian. This procedure leads to the following 
Hamiltonian:
\begin{eqnarray}
\label{H}
\hat{\cal H} = \sum_{\bf k} \varepsilon_{\bf k} b^\dagger_{\bf k}
b^{_{}}_{\bf k} +
\frac{1}{2} \sum_{{\bf k},{\bf q}} V_{{\bf k},{\bf q}} \left(
b^\dagger_{\bf k-q} b^\dagger_{\bf q} b^{_{}}_{\bf k}  
+ {\rm h.c.}\right) + ...\,, 
\end{eqnarray}
where the ellipses stand for the classical energy, other 3-boson
terms that do not lead
to decays, 4-boson, and the higher-order terms. 
Although we will need some of these other terms for the
$1/S$-expansion below,
Eq.~(\ref{H}) will suffice for the purpose of this paper. 
All of the necessary terms can be
found in \cite{miyake,chubukov94}.
The LSWT magnon energy and the 3-boson
vertex in Eq.~(\ref{H}) are given by:
\begin{eqnarray}
\label{E}
&& \varepsilon_{\bf k} =
  3JS\sqrt{(1-\gamma_{\bf k})(1+2\gamma_{\bf k})} \ ,\\
\label{V}
&& V_{{\bf k},{\bf q}} = 3iJ\sqrt{3S/2}\:\big[ 
f_{{\bf q},{\bf q'},{\bf k}}+f_{{\bf q'},{\bf q},{\bf k}}-
g_{{\bf k},{\bf q},{\bf q'}}\big]
\end{eqnarray}
where $J$ is the exchange constant,  ${\bf q'}\!=\!{\bf
  k}\!-\!{\bf q}$, $f_{1,2,3}=\bar\gamma_1(u_1+v_1)(u_2u_3+v_2v_3)$, 
$g_{1,2,3}=\bar\gamma_1(u_1+v_1)
(u_2v_3+v_2u_3)$,
$\gamma_{\bf k}
=\frac13(\cos k_x + 2\cos \frac{k_x}{2}\cos\frac{\sqrt{3}k_y}{2})$,
$\bar\gamma_{\bf k} =\frac13(\sin k_x - 2\sin
\frac{k_x}{2}\cos\frac{\sqrt{3}k_y}{2})$,   
and $u$, $v$ are the Bogolyubov coefficients: 
$2u_i^2-1=3JS(1+\frac{1}{2}\gamma_i)/\varepsilon_i$, $u_i^2-v_i^2=1$. 

\begin{figure}[t]
\centerline{
\includegraphics[width=0.6\columnwidth]{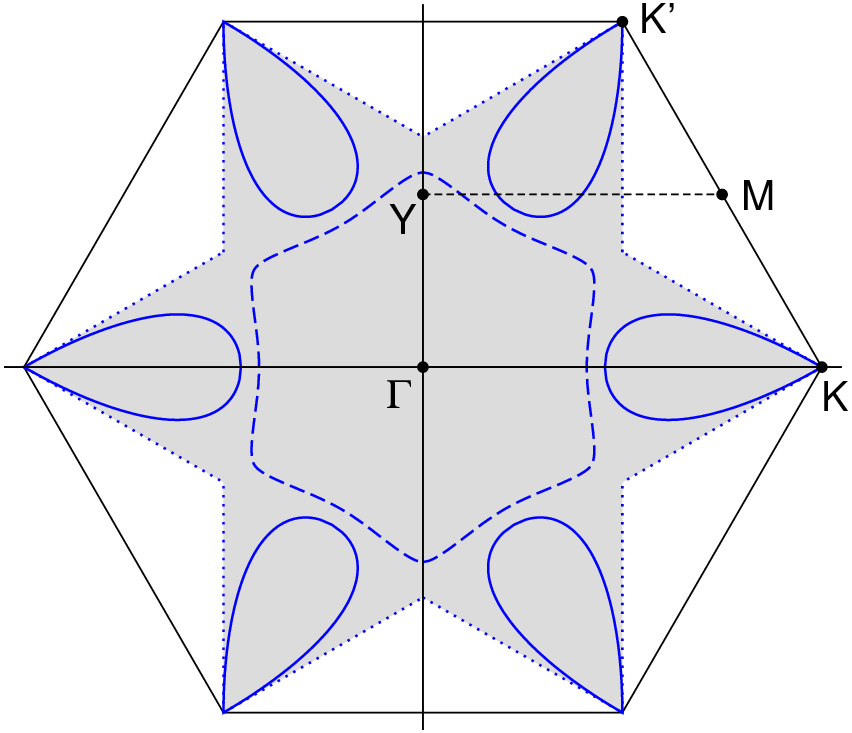}}
\caption{(color online). Brillouin zone of the TL.
The shaded area corresponds to the region
where spontaneous two-magnon decays are allowed. $\Gamma=(0,0)$, 
K=$(4\pi/3,0)$, K'=$(2\pi/3,2\pi/\sqrt{3})$, 
Y=$(0,\pi/\sqrt{3})$, and M=$(\pi,\pi/\sqrt{3})$ 
points are highlighted. The lines correspond to the extrema in the two-magnon
continuum described in the text.}
\label{BZ}
\end{figure}

{\it Kinematics: long wave-length limit.} ---
The magnon branch in Eq.~(\ref{E}) has three zero-energy modes:  
${\bf k}=0$ and ${\bf k}\!=\!\pm\!{\bf Q}\!=\!(\pm\frac{4}{3}\pi,0)$,
points $\Gamma$, and K (K') in Fig.~\ref{BZ}, respectively. 
In contrast with a square-lattice AF, velocities of 
these Goldstone modes are different: 
$c_{0}\!=\!\sqrt{3/4}$ and $c_{\pm Q}\!=\!\sqrt{3/8}$ (in units of $3JSa$).
This immediately implies that excitations with
${\bf k}\rightarrow 0$ are kinematically unstable towards decays into
$({\bf q}, {\bf q}')\rightarrow({\bf Q},-{\bf Q})$ 
ones, in close analogy with a 
decay of a longitudinal phonon into two transverse ones \cite{ziman}.
Clearly, such decays are immune to $1/S$ corrections as long as
velocities remain different.  
This picture is pertinent to all other non-collinear AF with
more than one Goldstone mode. 

For magnons at small $\tilde{\bf k}={\bf k}-{\bf Q}$ 
there exists a more subtle reason for decays. Instead of the usual 
convex and isotropic form, the magnon energy is nonanalytic
with varying convexity: 
$\varepsilon_{\bf k}\approx c_Q \tilde k(1-\alpha_\varphi\tilde k)$, where 
$\alpha_\varphi \sim \cos 3\varphi$.
This form together with the commensurability of the 
ordering vector create kinematic conditions for the decays 
from the steeper side of the energy cone at
${\bf k}\rightarrow{\bf Q}$ into the less
steeper sides at ${\bf q}, {\bf q}'\rightarrow-{\bf Q}$. 
Thus, magnons near the ${\bf Q}$-point are unstable only in a range 
of angles. Although such conditions are more delicate, 
it is very unlikely that the higher-order terms 
would selectively cancel the nonanalyticity.
Therefore, the ${\bf k}\rightarrow{\bf Q}$ decays 
should be prominent in the TLAF.

{\it Kinematics: full BZ.} ---
In the model (\ref{H}) an excitation with the momentum ${\bf k}$ is
unstable if the minimum energy of the two-particle continuum, 
$E_{{\bf k},{\bf q}}=\varepsilon_{\bf q} 
+\varepsilon_{{\bf k}-{\bf q}}$, is lower than $\varepsilon_{\bf k}$. 
Then the boundary between stable and unstable excitations is where 
such a minimum crosses the single-particle branch and the decay
condition $E_{{\bf k},{\bf q}}=\varepsilon_{\bf k}$ is first met.
Thus, to find these boundaries one should analyze 
the extrema of the continuum. 
For a gapless spectrum there can be several solutions 
as we show for the TLAF: \\
(a) Decay with emission of a ${\bf q}=0$ magnon. $E_{{\bf k},0} 
\equiv\varepsilon_{\bf k}$ for any ${\bf k}$ but it never
crosses the magnon branch. \\
(b) Decay with emission of ${\bf q}=\pm {\bf Q}$ magnons.
Equation that defines the boundary is 
$E_{{\bf k},{\bf Q}}\equiv\varepsilon_{{\bf k}\pm{\bf Q}} = 
\varepsilon_{\bf k}$ and its solution is shown by the dotted
line in Fig.~\ref{BZ}. The shaded area is where magnon decays are
allowed. It can be shown that $E_{{\bf k},{\bf Q}}$ corresponds to 
an absolute minimum of the continuum within the shaded area. 
In accord with our long wave-length discussion, the area
around ${\bf k}=0$ is enclosed and it is a finite segment in the 
vicinity of the ${\bf Q}$-point where decays are allowed. \\
(c) Decay into two identical magnons. 
The two-magnon continuum has extrema that are 
found from $\partial E_{{\bf k},{\bf q}}/\partial{\bf q}=0$ 
\cite{pitaevskii,field}. This condition means that the products
of decay have equal velocities. The simplest way to satisfy that
is to assume that their momenta are also equal. 
This is fulfilled automatically if ${\bf q}=({\bf k}+{\bf G}_i)/2$, 
where ${\bf G}_i$ is one of the two reciprocal lattice vectors of a TL, 
${\bf G}_1=(2\pi,2\pi/\sqrt{3})$ and ${\bf G}_2= 
(0,4\pi/\sqrt{3})$. The curves for the solution of 
$\varepsilon_{{\bf k}}=2\varepsilon_{({\bf k}+{\bf G}_i)/2}$  
 are shown in Fig.~\ref{BZ} by the solid lines. 
Note, that in the case of the TLAF these extrema 
are not the minima, but the saddle-points. Nevertheless, they 
lead to  essential singularities in the spectrum as will be discussed
below. \\
(d) Decay into non-identical magnons. In a more 
general situation, the decay products with the same velocities may 
have different momenta and energies $\varepsilon_{{\bf k}-{\bf q}} 
\neq \varepsilon_{\bf q}$.
Then, one has to solve the decay condition 
$E_{{\bf k},{\bf q}}=\varepsilon_{\bf k}$ together with the
extremum condition $\partial E_{{\bf k},{\bf q}}/\partial {\bf q}=0$. 
The solution is shown in
Fig.~\ref{BZ} by the dashed line. As in (c), 
corresponding extrema are the
saddle-points. 

Thus, the area of two-magnon decays in Fig.~\ref{BZ} is determined by the
solution (b) as it encloses regions
(c) and (d). This may not be the case for other systems (see below
the $XXZ$ model). 
Generally, the area of the decays is 
a union of the regions given by (b), (c), and (d).

{\it Spectrum: $1/S$ corrections.} ---
The $1/S$-correction to the TLAF magnon spectrum is given by the one-loop 
self-energy diagrams from the 3-boson terms, and the $\omega$-independent 
4-boson contribution, 
$\delta\varepsilon_{\bf k}=\Sigma_{\bf k}
(\varepsilon_{\bf k})+\delta\varepsilon^{(4)}_{\bf k}$:
\begin{eqnarray}
\label{S}
\Sigma_{\bf k}(\omega)&=&\frac{1}{2}\sum_{\bf q}\left(
\frac{|V_{{\bf k},{\bf q}}|^2}
{D^-_{\bf k}(\omega)} 
-\frac{|\widetilde{V}_{{\bf k},{\bf q}}|^2}
{D^+_{\bf k}(\omega)}\right)\\
\label{dE}
\delta\varepsilon^{(4)}_{\bf k}&=&
9J^2S\left[A_1\gamma_{\bf k}^2 +A_2\gamma_{\bf k} -A_3\right]
/4\varepsilon_{\bf k}
\end{eqnarray}
where $D^\mp_{\bf k}(\omega)=
\omega \mp \varepsilon_{\bf q} \mp \varepsilon_{\bf k-q}\pm i0$, 
the ``source''
3-boson vertex is ${\widetilde{V}}_{{\bf k},{\bf q}} = 
3iJ\sqrt{3S/2}\:\big[g_{\bf q,q',k}+g_{\bf q',q,k}-g_{\bf k,q,q'}\big]$, 
and $A_1=(4c_0\!+c_1\!-5c_2\!-4)$, $A_2=(-2c_0\!+c_1\!+c_2\!+2)$, $A_3=(
2c_0\!+2c_1\!-4c_2\!-2)$. Here, we have defined the following constants:
$c_n = 3JS \sum_{\bf k} \gamma_{\bf k}^n/\varepsilon_{\bf k}$,
$c_0 = 1.574733$,  $c_1 = -0.104254$, $c_2 = 0.344446$.

\begin{figure}[t]
\includegraphics[width=1.0\columnwidth]{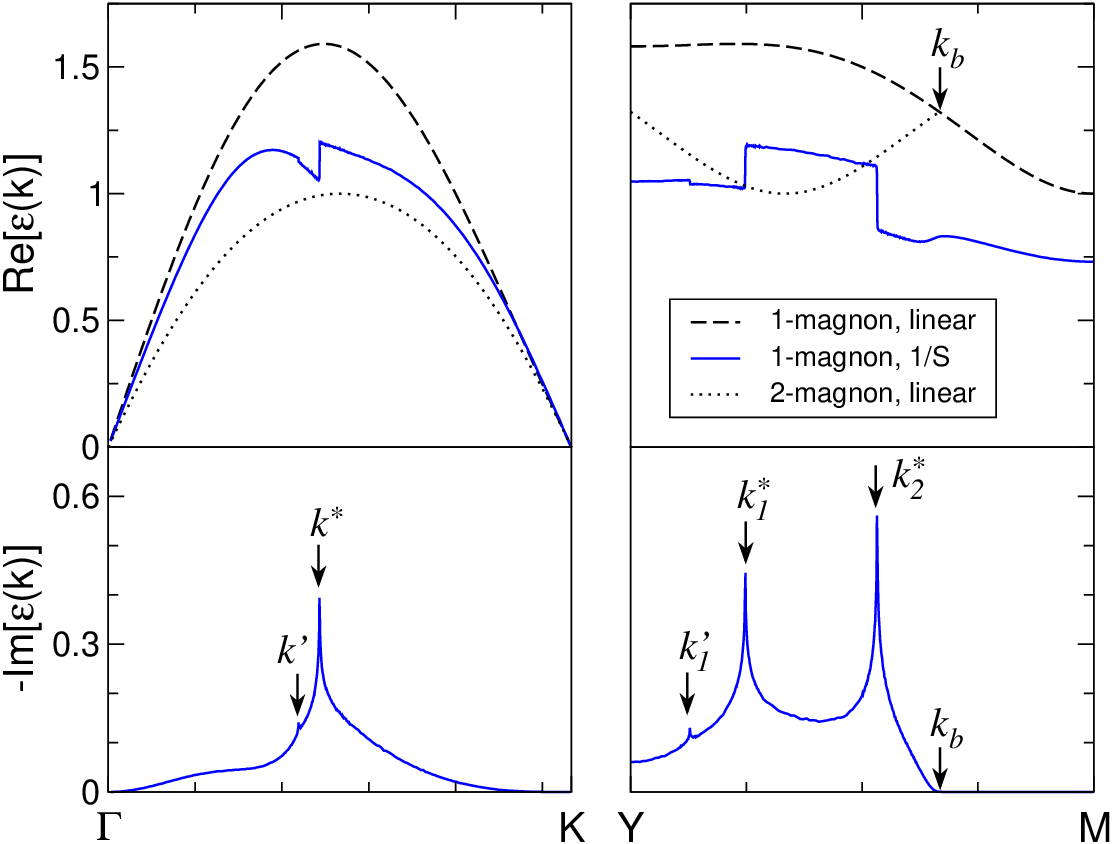}
\caption{(color online). Upper row: magnon dispersion along the
  lines $\Gamma$K and YM in the BZ, see Fig.~\ref{BZ}. 
The dashed lines are the LSWT spectrum, the dotted lines is the bottom
of the LSWT two-magnon continuum, and the solid lines are 
the spectrum with the $1/S$ correction. 
Lower row: imaginary part of the $1/S$ magnon energy 
along the same lines. $k_b$ is the intersection point of the one-magnon 
branch with the two-magnon continuum along the YM line. 
$k^*$ and $k'$ points correspond to the singularities discussed in text.}
\label{disp}
\end{figure}

We calculate the spectrum $\tilde\varepsilon_{\bf k}
=\varepsilon_{\bf k} 
+\delta\varepsilon_{\bf k}$ for $S=1/2$ 
using numerical integration in Eq.~(\ref{S}). 
Fig.~\ref{disp} shows $\tilde\varepsilon_{\bf k}$
for two representative directions in the BZ
together with 
the LSWT spectrum and the bottom of the two-magnon continuum.
The $1/S$-renormalization of the magnon spectrum is quite substantial,
see also \cite{Starykh}. This is because
the continuum strongly overlaps with, or,  for 
the ${\bf k}$-areas outside the decay region, has significant weight 
in a close vicinity of, the magnon branch. 
Such a purely kinematic effect explains a mystifying dichotomy: 
quantum corrections to the spectrum in the TLAF are large compared to the 
square lattice AF, while the ordered moments are about 
the same \cite{miyake,chubukov94,Singh}.

{\it Decays: long wave-length limit.} ---
The decay vertex (\ref{V}) for magnons near the $\Gamma$ point 
scales as 
$V_{\bf k,Q+q}\!\propto (q'-q)\!\sqrt{k/qq'}$, for small $q$ and  
$q'\!=\!|{\bf k-q}|$.
A simple power counting yields the leading term 
in the imaginary part of the self-energy. In a typical decay
$q, q'\!\sim\!k$ giving for the decay probability: 
$|V_{\bf k,Q+q}|^2\!\propto\!k$.  Since 
there is no constraint on the angle between ${\bf k}$ and
${\bf q}$, the 2D phase volume restricted by the
energy conservation contributes another factor of $k$,  such that
$\Im{\rm m}\{\Sigma_{\bf k}(\varepsilon_{\bf k})\}\propto
k^2$. A more detailed analytical calculation yields
$\Im{\rm m}\{\Sigma_{\bf k}(\varepsilon_{\bf k})\}\approx 
-0.789 Jk^2$, in agreement with the data in Fig.~\ref{disp}.

The decay vertex for ${\bf k}\!\rightarrow\!{\bf Q}$ magnon has
a more conventional scaling: $V_{\bf Q+k,-Q+q}\!\propto\! 
\sqrt{kqq'}$, so the decay
probability is $|V|^2\!\propto\!k^3$. Due to  
a constraint on the angle between ${\bf k}$ and ${\bf q}$, 
the decay surface in ${\bf q}$-space is a cigar-shaped ellipse 
with length $\sim\!k$ and width $\sim\!k^{3/2}$ that 
makes the restricted phase volume of decays to scale as $k^{1/2}$. 
This results in a nontrivial $k^{7/2}$
scaling of the decay rate. Numerically, along the $\Gamma$K line
$\Im{\rm m}\{\Sigma_{\bf k}(\varepsilon_{\bf k})\}\approx-1.2Jk^{7/2}$.
In a similar manner, one can show that at the boundary of 
decay region ({\it e.g.}, point $k_b$ in  Fig.~\ref{disp})
the decay rate grows as
$\Im{\rm m}\{\Sigma_{\bf k}(\varepsilon_{\bf k})\}\propto (k-k_b)^2$.

{\it Spectrum: singularities due to decays.} ---
A remarkable feature of the spectrum in Fig.~\ref{disp} is the 
singularities in the real and the imaginary parts of
$\tilde\varepsilon_{\bf k}$. Clearly, they are due to spontaneous
decays as it is only the decay term in (\ref{S}) 
that contributes to the imaginary part.

\begin{figure}[b]
\includegraphics[width=1.0\columnwidth]{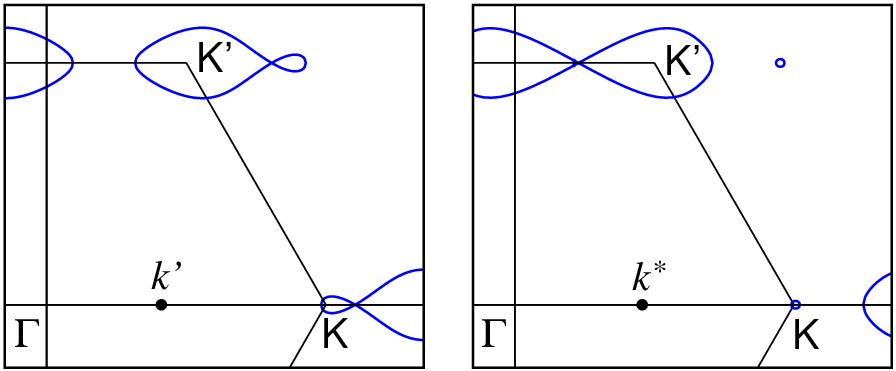}
\caption{(color online). The upper right parts of the BZ 
in the ${\bf q}$-space. The decay contours for $k=k'$
(left) and $k=k^*$ (right) along the $\Gamma$K line are shown. Both ${\bf
k}$-points correspond to the saddle points in the two-magnon
continuum. The corresponding ${\bf q}$-contours undergo topological
transition.}
\label{decay}
\end{figure}

A close inspection shows that $k'$ and $k^*$ singularity points 
in Fig.~\ref{disp} correspond exactly to the intersection of 
$\Gamma$K and YM lines with the saddle-points
in the continuum, solid and dashed lines in Fig.~\ref{BZ}.
The ``strong'' ($k^*$) and ``weak'' ($k'$) singularities
correspond to decays into identical and non-identical magnons, 
solutions (c) and (d) above, respectively. 
Fig.~\ref{decay} shows the decay contours, {\it i.e.}, 1D surfaces
in a 2D ${\bf q}$-space into which a magnon with the momentum 
${\bf k}$ can decay, for 
$k'$  and $k^*$  saddle points along the $\Gamma$K line. 
In both cases, these contours undergo topological
transition. 

Close to such a transition the denominator in Eq.~(\ref{S}) is expanded as
$\varepsilon_{\bf k}\!-E_{{\bf k},{\bf q}}\!\approx\!
(v_1\!-\!v_2)\Delta k\!-\!
\beta_x q_x^2\!+\!\beta_y q_y^2$,
$v_1$ and $v_2$ are velocities of the initial and final
magnons, respectively, $\beta_i$ are constants, $\Delta k\!=\!k\!-\!k^*$, 
$k^*$ is the saddle point. 
Integration in Eq.~(\ref{S}) yields a logarithmic singularity in the
imaginary part ${\Im}{\rm m}\{\Sigma_k\}\propto-\ln|\Lambda/\Delta k|$
and a concomitant finite jump in the real part of the self-energy
${\Re}{\rm e}\{\Sigma_k\}\propto {\rm sign}(\Delta k)$, 
$\Lambda$ is a cutoff.
The cutoff (size of the ``bubble'' in Fig.~\ref{decay}) 
is small in the case  of ``weak'' singularities,
which explains the weakness of them.
The logarithmic form agrees perfectly with the data in Fig.~\ref{disp}.
Overall, our analysis of the long- and short wave-length behavior 
gives complete understanding of the $1/S$ results for
the magnon spectrum in TLAF.

An important question is whether the singularities in the
spectrum will withstand the higher-order treatment.
The first possibility is when at least one of the final magnons 
is itself unstable. Then the log-singularity will be cut off
by the decay rate of the final magnon and ${\Im}{\rm m}\{\Sigma_k\}$ 
will have, at most, a weak maximum near the topological transition. 
For the TLAF such a scenario is realized for a large fraction of the ``weak''
singularities ($k'$ in Fig.~\ref{disp}). 
However, all of the ``strong'' singularities 
($k^*$ in Fig.~\ref{disp}) and some of the ``weak'' 
ones belong to another class, in which both magnons created in the
decay are stable. 
We have checked that the first-order $1/S$ corrections 
do not shift appreciably the instability  boundaries 
in Fig.~\ref{BZ}. Hence, at the saddle points, 
the logarithmic divergence of the one-loop diagrams will persist 
even for the renormalized spectrum.
In such a case, vertex corrections become important \cite{pitaevskii}.
Summation of an infinite  series of loop diagrams
yields the self-energy from the decay processes near the singular point:
$\Sigma_{\bf k}(\omega) \simeq ia/\ln|\Delta\omega-v_2\Delta k|$, with $a>0$.
One may conclude that the decay rate becomes vanishingly small
as $\Delta\omega,\Delta k\rightarrow 0$. This is, however, not true.
An attempt to solve the Dyson's equation $G_{\bf k}^{-1}(\omega)=0$ 
with this self-energy yields no solution for $\omega$ near the real axis. 
Therefore, the decay rate of quasiparticles around 
solid lines in Fig.~\ref{BZ} will remain large
and quasiparticle peaks will be strongly suppressed even
for large $S$.
\begin{figure}[t]
\includegraphics[width=1.0\columnwidth]{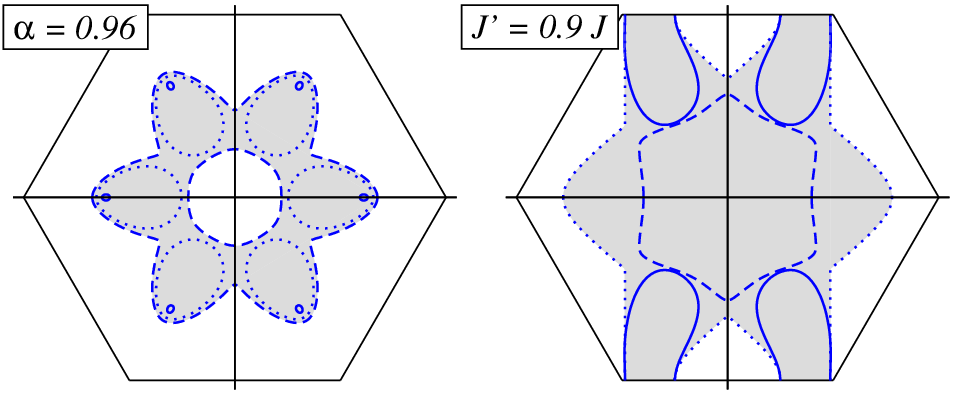}
\caption{(color online). Decay regions and singularity lines for 
the $XXZ$ model, $\alpha=0.96$ (left) and the 
$J-J'$ model, $J'/J=0.9$ (right). Definition of lines is the same as 
in Fig.~\ref{BZ}.}
\label{alpha}
\end{figure}

{\it Other models on a TL.} --- Two straightforward 
generalizations of the Heisenberg model on a TL are
(i) the anisotropic $XXZ$ model with $\alpha=J_z/J_{xy}<1$ 
and (ii) the $J$--$J'$ model for an orthorhombically distorted 
triangular lattice with one of the interactions within the triangle 
($J$) stronger than the other two ($J'$).

In the spectrum of the $XXZ$ model magnons at ${\bf Q}$ are gapped with 
the gap $\varepsilon_{\bf Q}\!\propto\!\sqrt{1-\alpha}$ 
at $\alpha\!\alt\!1$. This 
has two immediate consequences: (a) magnons at ${\bf k}\rightarrow 0$ 
are stable up to $\varepsilon_{\bf k}=2\varepsilon_{\bf Q}$ and
(b) ${\bf Q}$-magnons become stable themselves.
Thus, the star-shaped decay region in Fig.~\ref{BZ} develops a hole in 
the middle and have vertices shrunk and rounded. 
The evolution of the decay boundary with $\alpha<1$ is
non-trivial. Initially, the emission of a ${\bf Q}$ magnon remains 
an absolute minimum of the two-magnon continuum for most 
of the decay region. 
At $\alpha_1\!\approx\!0.993$ 
the decay into non-equivalent magnons switches from being a line of saddle 
points into the absolute minima of the continuum and takes over the
decay boundary.  
Fig.~\ref{alpha} shows the instability 
and the singularity lines for a  representative value $\alpha=0.96$. 
Further decrease of $\alpha$ completely eliminates the decay 
region at around $\alpha\approx 0.92$. 
Thus, magnon decays are present in an anisotropic TLAF, but only at not too 
large anisotropies.

For the $J$--$J'$ model the Goldstone modes at ${\bf k}=\pm{\bf Q}$ 
are preserved but the ordering wave-vector ${\bf Q}$ 
becomes incommensurate. This does not change the kinematics 
of the decays for the ${\bf k}\rightarrow 0$  magnons, 
but forbids the decays from the vicinity of the ${\bf Q}$ point into 
the vicinity of $-{\bf Q}$ point as the quasi-momentum cannot be conserved. 
However, the decays in the vicinity of (inequivalent now) K' points are still 
allowed. Overall, the decay region grows with the decrease of $J'$. At 
$J'\approx 0.34J$, relevant to Cs$_2$CuCl$_4$ \cite{Coldea}, 
the decay region covers most of 
the BZ. With the decrease of $J'$ the LSWT single-magnon dispersion develops 
a low-energy branch in the direction perpendicular 
to the ``strong'' $J$. That makes the rest of the spectrum prone to
decays into it. 

{\it Conclusions.} --- We have 
shown that magnon decays must be prominent in a wide class of
noncollinear AFs. We calculated the decay rate in the spin-1/2 TLAF
within the spin-wave theory.
In the long-wavelength limit, the life-time of low-energy excitations is
predicted to exhibit a non-trivial scaling. For the short-wavelength
magnons,  the decay rate is large, $2\Im{\rm
  m}\{\tilde\varepsilon_{\bf k}\} \sim  
0.4\Re{\rm e}\{\tilde\varepsilon_{\bf k}\}$, in a substantial part
of the BZ. Topological transitions of the decay surface also lead
to strong singularities in the spectrum that remain essential even for large
values of spin. Therefore, excitations in ordered, spin-$S$, AFs may not
necessarily be well-defined for all wave-vectors.

{\it Acknowledgments.} ---
We are indebted to O. Starykh for illuminating
discussions and sharing his unpublished work.
This work was supported by DOE under grant DE-FG02-04ER46174 (A.L.C.).


\begin{thebibliography}{99}

\bibitem{ziman}
J. M. Ziman, {\it Electrons and phonons}, (Oxford University Press,
Oxford, 1960).

\bibitem{pitaevskii}
L. P. Pitaevskii, Zh. \'Eksp. Teor. Fiz. {\bf 36}, 1168 (1959)
[Sov. Phys. JETP {\bf 9}, 830 (1959)].

\bibitem{collinear} F. J. Dyson, 
 Phys. Rev. {\bf 102}, 1217 (1956); A. B. Harris {\it et al}.,
Phys. Rev. {\bf 3}, 961 (1971).

\bibitem{akhiezer}
A. I. Akhiezer, V. G. Bar'yakhtar, S. V. Peletminskii,
{\it Spin waves} (North-Holland, Amsterdam, 1968).

\bibitem{miyake}
S. J. Miyake, Prog. Theor. Phys. {\bf 73}, 18 (1985);
J. Phys. Soc. Jpn. {\bf 61}, 983 (1992).

\bibitem{ohyama}
T. Ohyama and H. Shiba, 
J. Phys. Soc. Jpn. {\bf 62}, 3277 (1993).

\bibitem{chubukov94}
A. V. Chubukov {\it et al}., 
J. Phys. Condens. Matter {\bf 6}, 8891 (1994).

\bibitem{Zh_nikuni} M. E. Zhitomirsky and T. Nikuni,
 Phys. Rev. B {\bf 57}, 5013 (1998).

\bibitem{field}
M. E. Zhitomirsky and A. L. Chernyshev,
Phys. Rev. Lett. {\bf 82}, 4536 (1999).

\bibitem{Singh}
W. Zheng {\it et al}.,
Phys. Rev. Lett. {\bf 96}, 057201 (2006).

\bibitem{Starykh}
O. A. Starykh {\it et al}.,
Phys. Rev. B {\bf 74}, 180403(R) (2006).

\bibitem{Coldea}
R. Coldea {\it et al}.,
Phys. Rev. Lett. 86, 1335 (2001).


\end{thebibliography}
\end{document}